\documentclass[11pt,a4paper]{article}

\usepackage{fullpage}
\usepackage{indentfirst}
\usepackage[utf8x]{inputenc}
\usepackage[T1]{fontenc}
\usepackage{hyperref}
\usepackage[english]{babel}
\usepackage{amsmath}
\usepackage{amsfonts}
\usepackage{amssymb}
\usepackage{amscd}
\usepackage{graphicx}
\usepackage{pstricks}
\author{D. Schuch$^{1}$, J. Guerrero$^{2}$, F.F. L\'{o}pez-Ruiz$^{3}$ and V. Aldaya$^{4}$}

\title{Interrelations between different canonical descriptions of dissipative systems}

\date{\begin{center}
\begin{small}$^1$ Institut f\"ur Theoretische Physik, J.W. Goethe-Universit\"at Frankfurt am Main, 
\end{small}\\	
\begin{small}
  Max-von-Laue-Str. 1, D-60438 Frankfurt am Main, Germany
\end{small}\\
\begin{small}$^2$Departamento de Matem\'{a}tica Aplicada, Universidad de Murcia, 
\end{small}\\
\begin{small}Campus de Espinado, E-30100 Murcia, Spain
\end{small}\\
\begin{small}$^3$Departamento de F\'{i}sica Aplicada, Universidad de C\'{a}diz, 
\end{small}\\
\begin{small}Campus de Puerto Real, E-11510 Puerto Real, C\'{a}diz, Spain
 \end{small}\\
\begin{small}$^4$Instituto de Astrof\'{i}sica de Andaluc\'{i}a, IAA-CSIC,
\end{small}\\
\begin{small} Apartado Postal 3004, E-18080 Granada, Spain
\end{small}\\
\begin{small}Schuch@em.uni-frankfurt.de \ juguerre@um.es \ paco.lopezruiz@uca.es \ valdaya@iaa.es
\end{small}\\                                                                   
\end{center}
}
\begin{document}

\maketitle
%


\begin{abstract}
There are many approaches for the description of dissipative systems coupled to some kind of environment. This environment can be described in different ways; only effective models will be considered here. In the Bateman model, the environment is represented by one additional degree of freedom and the corresponding momentum. In two other canonical approaches, no environmental degree of freedom appears explicitly but the canonical variables are connected with the physical ones via non-canonical transformations. The link between the Bateman approach and those without additional variables is achieved via comparison with a canonical approach using expanding coordinates since, in this case, both Hamiltonians are constants of motion. This leads to constraints that allow for the elimination of the additional degree of freedom in the Bateman approach. These constraints are not unique. Several choices are studied explicitly and the consequences for the physical interpretation of the additional variable in the 
Bateman model are discussed.
\end{abstract}

\section{Introduction}

Realistic physical systems are not isolated but in contact with some kind of
environment causing phenomena like irreversibility of the time-evolution and
dissipation of energy. These kinds of effects can be described by
phenomenological equations of motion like the Langevin equation with a linear
velocity dependent friction force. But this does not fit into the conventional
Lagrangian or Hamiltonian formalism of classical mechanics where the canonical
variables are the physical position and momentum or related with them via
canonical transformations and the Hamiltonian function is the sum of kinetic
and potential energies. Attempts to obtain the afore-mentioned friction force
by adding a kind of ``friction potential'' to the Hamiltonian have not been
successful (at least not on the classical level). However, other attempts to
include friction effects into the Hamiltonian formalism exist but different
prices have to be paid for this purpose.

In the conventional system-plus-reservoir approach, the system of interest is
coupled to an environment with many (in the limit infinitely many) degrees of
freedom (e.g. linearly coupled to a bath of harmonic oscillators \cite{1})
where the system and the environment together are considered to be a closed
Hamiltonian system. Via averaging over the environmental degrees of freedom
and other procedures (for details see, e.g., \cite{2}), an equation of motion
for the system of interest including a friction force can finally be
obtained. One drawback in employing this method is the large number of
environmental degrees of freedom that must be considered in the beginning
(though they are eliminated in the end). This leads to large, cumbersome and
expensive calculations. In its quantized version, this approach is usually
applied to the density matrix causing the computational effort to scale at
least quadratically with the number of degrees of freedom and, in the case of
the Caldeira--Leggett model \cite{1}, can also lead to unphysical negative probabilities.

The number of environmental degrees of freedom is drastically reduced to one
in an approach by Bateman \cite{3} to describe the damped harmonic
oscillator. In order to be able to apply the canonical formalism, the
phase-space dimension must be doubled to obtain a kind of effective
description. The new degree of freedom can be considered as a collective one
for the bath that absorbs energy dissipated by the damped oscillator. The
variable of the dual system that fulfills a time-reversed equation with an
acceleration force of the same magnitude as the friction force of the Langevin
equation, but with a different sign, looks like a position variable and its
relation to, and interpretation in terms of, physical position and momentum
(or velocity), particularly when linked to other canonical approaches, will be investigated in this work.

After the rediscovery of the Bateman dual Hamiltonian by Morse and Feshbach
\cite{4} and Bopp \cite{5} it has been studied with respect to various
different features also in recent years. So squeezed states for the Bateman
Hamiltonian were considered in \cite{6} and \cite{7} and a quantum field
theoretical approach was used by Vitiello et al \cite{8}. This author also
tried to apply the dual approach as a dissipative quantum model of the brain
\cite{9}. Quantization using Feynman's path integral method was
discussed by Blasone and Jizba \cite{10,11} and the Bateman system has also
been studied by the same authors and Vitiello \cite{11,12} as a toy model for
't Hooft's proposal of a deterministic version of quantum mechanics
\cite{13}. More recently, together with Scardigli, these authors considered a
composite system of two classical Bateman oscillators as a particle in an
effective magnetic field \cite{14}. Complex eigenvalues of the quantized
version of Bateman's Hamiltonian in connection with resonances and
two-dimensional parabolic potential barriers are discussed in
\cite{15,16}. Also, the Wigner function for the Bateman system on
non-commutative phase space \cite{17} and the inclusion of a time-dependent
external force \cite{18} have been studied. The Bateman approach (as well as
the one of Caldirola \cite{19} and Kanai \cite{20} that will be considered
subsequently) is also discussed in an attempt to reformulate a dissipative
system in terms of an infinite number of non-dissipative ones \cite{21}. A
different method for the description of dissipative systems that seems to have
some advantages in the high energy regime has been compared with the Bateman
approach \cite{22} and shown to be locally equivalent to it. Finally, a
rather recent paper \cite{23} by Bender et al studies the Bateman Hamiltonian
enlarged by a quadratic term in the two dual coordinates as a model for two
coupled optical resonators. This shows that, despite the age of Bateman's
approach, there is still considerable interest in, and potential applicability
of, this model. 

Another frequently applied approach for the description of dissipative systems
that does not take into account the individual degrees of freedom of the
environment is the one of Caldirola \cite{19} and Kanai \cite{20}. Actually no
environmental degree of freedom appears expilcitly in this
approach, only the effect of the environment on the system of interest is taken into
account. This method is a formal canonical one that again leads to an equation
of motion with the same damping force, but now derived from a Lagrangian or
Hamiltonian that contains no additional friction terms only a multiplying
factor. The corresponding Hamiltonian, however, no longer represents the
energy of the system and is also not a constant of motion. The most serious
point of criticism usually raised against this approach is its apparent
violation of the uncertainty principle in its quantized form that can be
obtained via canonical quantization. This criticism can be refuted if the
relation between the canonical variables, and quantities depending on them,
and the usual physical variables is properly taken into account (for details
see \cite{24}). In the following, however, only the classical version will be considered.

The final approach that will be mentioned in more detail here uses an
exponentially-expanding coordinate system \cite{25} \cite{26}. The canonical
position and momentum variables of this approach, as in the case of
Caldirola--Kanai (CK), are connected with the physical position and momentum
via a non-canonical transformation. In this case, however, the Hamiltonian is
a constant of motion and, for certain initial conditions, even represents the initial energy of the physical system
that is dissipated during its time-evolution. It formally looks like that of
an undamped harmonic oscillator, only with shifted frequency. Therefore, all
known results from the undamped oscillator can be used and the corresponding
results for the damped case are obtained via the non-canonical transformation
from the canonical to the physical system. In particular, after canonical quantization, no
problems occur with the uncertainty principle \cite{26}. This approach is
connected with the one of CK via a canonical transformation \cite{24} \cite{27} (however, with explicitly time-dependent generating function).

There are further similar canonical approaches using modified Lagrange and Hamilton functions for the system of interest, like the one by Lemos \cite{28} that also has a conserved Hamiltonian. But these approaches are related to the one in expanding coordinates (and therefore also with the one of CK) via canonical transformations and will not be considered further in this work (for details see also \cite{29}).

It has been shown by Sun and Yu \cite{30} \cite{31} that it is possible to get
to the CK Hamiltonian from the Caldeira--Leggett model thus demonstrating a
kind of physical equivalence of the two methods. On the other hand, group
theoretical arguments have been used to link the CK approach to the one by
Bateman \cite{32}. In this paper it will be shown explicitly how the Bateman
approach can be related to the canonical one using expanding coordinates. For
this purpose the variables of the dual system must be eliminated by imposing
some constraints; it will be shown how this can be expressed in terms of
physical position and velocity of the damped system. We will make use of the
circumstance that both Hamiltonians are constants of motion. The transition to
the CK system is then achieved simply via a time-dependent canonical transformation.

The discussion will be restricted to a one-dimensional system, in particular
the damped harmonic oscillator (where the damped free motion can be obtained
in the limit $\omega \rightarrow 0$) and to the classical case.

Following an outline of the Bateman model, there will be a short presentation
of the approach using the expanding coordinates and the one of CK as well as their interrelation. To find the connection with the Bateman approach the variables of the dual system will be removed by imposing constraints, which can be done in different ways. Some examples and their consequences will be discussed in detail and conclusions will be drawn at the end.

\section{The Bateman approach}

The Bateman Hamiltonian $H_B$, expressed in terms of the position variables
$x$ and $y$ and the corresponding canonical momenta $p_x$ and $p_y$, reads   
\begin{equation}
H_B~=~\frac{1}{m}~ p_x p_y~+~\frac{\gamma}{2}~ (y p_y~-~x p_x)~+~m \left(
  \omega^2 - \frac{\gamma^2}{4} \right) x y ~=~H_{\Omega}~+~D~,
\label{gran1}
\end{equation}
with $D = \frac{\gamma}{2}~ (y p_y~-~x p_x)$. The Poisson brackets of $H_B$
with $D$ as well as with $H_{\Omega}$ vanish, so both are constants of motion
(in the quantized version, the corresponding three operators commute). 

The Hamiltonian equations of motion are 
\begin{equation}
\frac{\partial H_B}{\partial p_x}~=~\frac{1}{m} p_y~-~\frac{\gamma}{2}
x~=~\dot{x}~,~~~~\frac{\partial H_B}{\partial p_y}~=~\frac{1}{m} p_x~+~\frac{\gamma}{2}
y~=~\dot{y}~,
\label{gran2}
\end{equation}
\begin{equation}
\frac{\partial H_B}{\partial x}~=~-~\frac{\gamma}{2}~p_x~+~m \left(
  \omega^2 - \frac{\gamma^2}{4} \right) y~=~-~\dot{p}_x~,~~\frac{\partial H_B}{\partial y}~=~\frac{\gamma}{2}~p_y~+~m \left(
  \omega^2 - \frac{\gamma^2}{4} \right) x~=~-~\dot{p}_y~,
\label{gran3}
\end{equation}
where, from (2), $p_x$ and $p_y$ can be expressed as
\begin{equation}
p_y~=~m~\left( \dot{x}~+~\frac{\gamma}{2} x \right)~,
\label{gran4}
\end{equation}
\begin{equation}
p_x~=~m~\left( \dot{y}~-~\frac{\gamma}{2} y \right)~,
\label{gran5}
\end{equation}

From there, and with the help of Eqs. (3), the equations of motion for $x$ and $y$ can be obtained as
\begin{equation}
\ddot{x}~+~\gamma \dot{x}~+~\omega^2 x~=~0~,
\label{gran6}
\end{equation}
\begin{equation}
\ddot{y}~-~\gamma \dot{y}~+~\omega^2 y~=~0~.
\label{gran7}
\end{equation}

Equation (6) is just the equation for the damped harmonic oscillator with
friction force $- \gamma \dot{x}$, whereas, in the time-reversed equation for $y$, the accelerating force $+ \gamma \dot{x}$ occurs.

From Eqs. (6), (7) and (4), (5) it is clear that the $(x, p_x, y, p_y)$ space
splits into two invariant subspaces: the one of variables $(x, p_y)$
undergoing a damped oscillator motion, and the one of variables $(y, p_x)$
with time-reversed (accelerated) behavior.

Using the equations of motion, it can also be shown that
\begin{equation}
\frac{d}{dt} H_B~=~0~,
\label{gran8}
\end{equation}
i.e., $H_B$ is a dynamical invariant which, in a first naive attempt, could be interpreted in the way that the energy dissipated by the damped system is gained by the accelerated one.
Rewritten in terms of $x$, $y$ and the corresponding velocities $\dot{x}$ and
$\dot{y}$, the terms depending on the friction (or acceleration) coefficient
$\gamma$ cancel out (although the Lagrangian does contain terms in $\gamma$) and it remains
\begin{equation}
H_B~\hat{=}~m (\dot{x} \dot{y}~+~\omega^2~xy~)~.
\label{gran9}
\end{equation}

In fact, the individual energies, and their change in time for both systems, written in terms of the velocities take the form
\begin{equation}
E_x~=~\frac{m}{2} \dot{x}^2~+~\frac{m}{2} \omega^2 x^2
~,
\label{gran10}
\end{equation}
with
\begin{equation}
\frac{d}{dt} E_x~=~- \gamma m \dot{x}^2 
~,
\label{gran11}
\
\end{equation}
and
\begin{equation}
E_y~=~\frac{m}{2} \dot{y}^2~+~\frac{m}{2} \omega^2 y^2  
~,
\label{gran12}
\end{equation}
with
\begin{equation}
\frac{d}{dt} E_y~=~+ \gamma m \dot{y}^2~.
\label{gran13}
\end{equation}

So, the sum of $E_x$ and $E_y$ would be constant and (apart from another
constant term) could be equal to $H_B$ if
\begin{equation}
\frac{d}{dt} (E_x~+~E_y)~=~\gamma~m (\dot{y}^2~-~\dot{x}^2)~=~0~,
\label{gran14}
\end{equation}
which is fulfilled only for $\dot{y} = \pm \dot{x}$; so $y$ and $x$ could differ, at most, by a
constant and $H_B$, as given in (9), (again apart from a constant term) would
turn into $H_B~\rightarrow~m (\dot{x}^2~+~\omega^2~x^2)$, i.e., the energy of two undamped harmonic oscillators.

However, $\dot{y}$, derived from the solution of Eq. (7), differs from
$\dot{x}$, derived from the solution of Eq. (6) by more than just its sign; so one has
to be careful with this simple picture of energy transfer between the x- and
y- systems. This is rather clear if we notice that both degrees of freedom in
the Bateman system (regardless of their physical interpretation) are so
involved that $H_B$ is not of the form $H_B = H_x + H_y + H_{xy}$, where $H_x$ and $H_y$ are
harmonic oscillator Hamiltonians for $x$ and $y$, and $H_{xy}$ is an interaction
term. One can rotate the phase space in order to obtain a new system of two
oscillators with opposite signs coupled through an interaction term (see, e.g.,
\cite{14}, Eqs. (21) - (23)) where again the energy of the whole system is
conserved and it is transferred from one of the transformed oscillators to the
other, but none of the coordinates of the rotated oscillators represents the
physical position variable. When constraints that are imposed on the systems are considered it becomes even more obvious later on that $y$ is not just a simple position coordinate like $x$ in this model.

\section{Effective canonical description of dissipative systems in expanding
  coordinates and in the CK-approach}

Now, briefly, two approaches are presented that are able to describe the
damped harmonic oscillator in the framework of Hamiltonian mechanics using
only one canonical position and momentum variable. These variables, however,
are connected with physical position and momentum via {\it non-canonical}
transformations. In the following, {\it canonical} variables and corresponding
Hamiltonians will be characterized by a hat.

\subsection{ Exponentially expanding coordinate system}

The Hamiltonian $\hat{H}_{exp}$ depends on a coordinate $\hat{Q}$ that, in
comparison with the physical position variable $x$, expands exponentially and
the corresponding canonical momentum $\hat{P}$ displays a similar behavior, i.e.,
\begin{equation}
\hat{H}_{exp}~=~\frac{1}{2m}~\hat{P}^2~+~\frac{m}{2} \left(
  \omega^2 - \frac{\gamma^2}{4} \right) \hat{Q}^2 
 \label{gran15}
\end{equation}
with
\begin{eqnarray}
\hat{Q}~=~x~e^{\frac{\gamma}{2}t}~~&,&~~ \hat{P}~=~m~ \dot{\hat{Q}}~=~m \left(\dot{x} +
\frac{\gamma}{2} x \right) e^{\frac{\gamma}{2}t}~, \nonumber\\
\textrm{and}~~~~~~~~~~~~~~ \frac{\partial \hat{H}_{exp}}{\partial \hat{P}}~=~\frac{1}{m} \hat{P}~=~\dot{\hat{Q}}~~&,&~~\frac{\partial \hat{H}_{exp}}{\partial \hat{Q}}~=~m \Omega^2
\hat{Q}~=~-~\dot{\hat{P}}
~.
\label{gran16}
\end{eqnarray}

Hamiltonian (15) looks like that of an undamped harmonic oscillator with
shifted frequency $\Omega = ( \omega^2 - \gamma^2/4 )^{1/2}$ and the
corresponding equation of motion for $\hat{Q}$ is consequently
\begin{equation}
\ddot{\hat{Q}}~+~\Omega^2 ~\hat{Q}~=~0~.
\label{gran17}
\end{equation}

Expressed in terms of the physical position variable $x$, Eq. (6) is regained, including the friction force.
Obviously, also 
\begin{equation}
\frac{d}{dt} \hat{H}_{exp}~=~0
\label{gran18}
\end{equation}
is valid which, too, can be confirmed by rewriting $\hat{H}_{exp}$ in terms of $x$ and $\dot{x}$ as
\begin{equation}
\hat{H}_{exp}~\hat{=}~\frac{m}{2} \left[ \dot{x}^2~+~\gamma \dot{x} x~+~ \omega^2
x^2 \right] e^{\gamma t}~=~\textrm{const}.~,
\label{gran19}
\end{equation}
which for $x_0 = 0$ or $\dot{x}_0 = 0$ even represents the initial energy of
the system. It is interesting to note that in general $\hat{H}_{exp}$, written
in terms of $x$ and $\dot{x}$, coincides with a conserved quantity for the
damped harmonic oscillator already considered in the literature (see, for
instance, the expression for $I_5$ in Eq. (19) of \cite{33}). In the context of
the Caldirola--Kanai description presented below, the eigenstates of the
quantum operator corresponding to this invariant $\hat{H}_{exp}$ are known as
loss-energy states (see \cite{34}), although the fact that $\hat{H}_{exp}$ is
constant allows to find the quantum operator that represents it in a broader context.

\subsection{ Caldirola--Kanai approach}

In the Caldirola--Kanai approach, the position variable remains unchanged,
$\hat{x} = x$, whereas only the canonical momentum shows an exponential expansion, i.e.,
\begin{equation}
\hat{H}_{CK}~=~\frac{1}{2m}~\hat{p}^2~ e^{-\gamma t}~+~\frac{m}{2} \omega^2
\hat{x}^2~e^{\gamma t}
\label{gran20}
\end{equation}
with
\begin{equation}
\hat{x}~=~x~~,~~\hat{p}~=~p~e^{\gamma t}~=~m~\dot{x}~~e^{\gamma t}~.
\label{gran21}
\end{equation}

From the Hamiltonian equations of motion
\begin{equation}
\frac{\partial \hat{H}_{CK}}{\partial \hat{p}}~=~\frac{1}{m} \hat{p}~e^{- \gamma
  t}~=~\dot{x}~~,~~\frac{\partial \hat{H}_{CK}}{\partial \hat{x}}~=~m \omega^2
\hat{x}~e^{\gamma t}~=~-~\dot{\hat{p}}
\label{gran22}
\end{equation}
it again follows that the physical position variable obeys Eq. (6). Expressed
in terms of $x$ and $\dot{x}$, $\hat{H}_{CK}$ now reads
\begin{equation}
\hat{H}_{CK}~\hat{=}~\frac{m}{2} \left[ \dot{x}^2~+~ \omega^2
x^2 \right] e^{\gamma t}~=~E(t)~ e^{\gamma t}~\neq~\textrm{const}.
\label{gran23}
\end{equation}

The canonical variables of this approach are connected with the ones of $\hat{H}_{exp}$ via
\begin{equation}
\hat{x}~=~\hat{Q}~e^{-\frac{\gamma}{2}t}~~\textrm{or}~~\hat{Q}~=~\hat{x}~e^{\frac{\gamma}{2}t}~,
\label{gran24}
\end{equation}
\begin{equation}
\hat{p}~=~\hat{P}~e^{\frac{\gamma}{2}t}~-~m~\frac{\gamma}{2}~\hat{Q}~e^{\frac{\gamma}{2}t}~~\textrm{or}~~\hat{P}~=~\hat{p}~e^{-
  \frac{\gamma}{2}t}~+~m \frac{\gamma}{2}~\hat{x}~e^{\frac{\gamma}{2}t}~,
\label{gran25}
\end{equation}
where the explicitly time-dependent generating function for the canonical transformation between the two systems is given by
\begin{equation}
\hat{F}_2(\hat{x},\hat{P},t)~=~\hat{x}~ \hat{P}~
e^{\frac{\gamma}{2}t}~-~m~\frac{\gamma}{4}~ \hat{x}^2~e^{\gamma t}~. 
\label{gran26}
\end{equation}

\section{ Linking the Bateman approach with $\hat{H}_ {exp}$}

In order to connect the Bateman approach with the canonical approaches presented in Section 3, it will be stipulated that

1) the equation of motion (6) for the position variable of the dissipative system is the same as the equation of motion for the position variable (including the friction force) in the two canonical approaches when these are expressed in terms of the physical position variable $x$.

2) The Bateman Hamiltonian represents a constant of motion with the dimension
of an energy. 

In order to connect the two descriptions of a dissipative system we may assume
that the conserved quantity $H_B$ is identical to the conserved quantity $\hat{H}_{exp}$ and impose some constraints so that the dual variable $y$ and the
corresponding momentum are eliminated.

Since the constraints are obtained via comparison with Hamiltonians on
the formal canonical level, the notation in $H_B$ will be changed and
variables with a hat will be used to distinguish them from the ones in the
original Hamiltonian. So, the Bateman Hamiltonian is now written as
\begin{equation}
\hat{H}_B~=~\frac{1}{m}~ \hat{p}_x \hat{p}_y~+~\frac{\gamma}{2}~ (\hat{y} \hat{p}_y~-~\hat{x} \hat{p}_x)~+~m \left(
  \omega^2 - \frac{\gamma^2}{4} \right) \hat{x} \hat{y}~,
\label{gran27}
\end{equation}
which must be compared with $\hat{H}_{exp}$, as given in (19), where $\hat{x}
= x$ is valid since $x$ fulfills Eq. (6) for the physical position
variable. From Eq. (4), it follows that $\hat{p}_y~=~m~\left(
  \dot{\hat{x}}~+~\frac{\gamma}{2} \hat{x} \right)~=~m~\left(
  \dot{x}~+~\frac{\gamma}{2} x \right)$ so none of the product terms of
$\hat{x}$ and $\hat{p}_y$ with one of the other variables $\hat{y}$ and
$\hat{p}_x$ in (27) contains the exponential factor $e^{\gamma t}$ that is
common in Eq. (19). 

Following the prescription outlined above, we equate  $\hat{H}_B$ (in the form
of Eq. (9)) with $\hat{H}_{exp}$:
\begin{equation}
\frac{m}{2}~e^{\gamma t}~\left[ \dot{x}^2~+~\gamma \dot{x} x~+~ \omega^2
x^2 \right]~=~\hat{p}_x \dot{x}~+~m \frac{\gamma}{2}~ \hat{y} \dot{x}~+~m
\omega^2 x  \hat{y}~,
\label{gran28}
\end{equation}
which is only possible if $\hat{y}$ and $\hat{p}_x$ are expressed in terms of $x$ and
$\dot{x}$. For this purpose the ansatz 
\begin{equation}
\hat{p}_x~=~e^{\gamma t}~\left(a~\dot{x}~+~b~x \right)~~~ \textrm{and}
~~~\hat{y}~=~e^{\gamma t}~\left(c~\dot{x}~+~d~x \right)~
\label{gran29}
\end{equation}
is inserted into (28) and the coefficients of $\dot{x}^2$-, $x \dot{x}$- and $x^2$-terms are equated leading to 
\begin{equation}
d~=~\frac{1}{2}~,
\label{gran30}
\end{equation}
\begin{equation}
a~=~\frac{m}{2}~(1~-~\gamma c)~,
\label{gran31}
\end{equation}
\begin{equation}
b~=~m~\left( \frac{\gamma}{4}~-~\omega^2~c \right)~,
\label{gran32}
\end{equation}
where $a$, $b$ and $c$ still have to be determined. Since only two equations
(31, 32) are given, one parameter is still free to be chosen. Note that
expressing (29) in terms of canonical variables,
\begin{eqnarray}
\hat{y}~=~e^{\gamma t}~\left(c~\dot{x}~+~d~x \right)~=~e^{\gamma
  t}~\left(\frac{c}{m}~\hat{p}_y~+~\left(d - \frac{\gamma}{2} c \right) x \right)~,   \nonumber \\
\hat{p}_x~=~e^{\gamma t}~\left(a~\dot{x}~+~b~x \right)~=~e^{\gamma
  t}~\left(\frac{a}{m}~\hat{p}_y~+~\left(b - \frac{\gamma}{2} a \right) x \right)~,
\label{gran33}
\end{eqnarray}
an explicit time-dependent character of the constraints shows up, although they are
compatible with the equations of motion, that is, the total time derivative of
the constraints is zero. The explicit dependence on time of the constraints is
traced back to the fact that they have non-vanishing Poisson brackets with the Hamiltonian.

In the following, a more detailed discussion of the cases 1) $c = 0$, 2) $a =
0$ and 3) $b = 0$ will be given.

Generally, one parameter can be eliminated leaving a condition for the
relation of, e.g., $a$ and $b$ that has to be fulfilled for any choice of $c$ etc.,
\begin{equation}
a~=~\frac{1}{2 \omega^2}~(m ~\Omega^2~+~\gamma~
b)~~\textrm{or}~~b~=~\frac{2}{\gamma}~(\omega^2~a~-~\frac{m}{2}~\Omega^2)~~.
\label{gran34}
\end{equation}

\subsection{ The case $c$ = 0}

For this choice of $c$ it follows that $a~=~\frac{m}{2}$ and
$b~=~m~\frac{\gamma}{4}$. In the following discussions, from the Bateman
system $\hat{x}~=~x~=~\hat{Q}~e^{-\frac{\gamma}{2}t}~$ and
$\hat{p}_y~=~m~(\dot{x} + \frac{\gamma}{2} x )~=~\hat{P}~e^{-\frac{\gamma}{2}t}$
will always be valid, only $\hat{y}$ and $\hat{p}_x$ expressed in terms of of $x$
and $\dot{x}$ will change. Therefore, $\hat{y}$ and $\hat{p}_x$ will be supplied with a
second subscript indicating which parameter has been set equal to zero.    

So, in this case, one obtains
\begin{equation}
\hat{y}_c~=~\frac{1}{2}~x~~e^{\gamma t}~=~\frac{1}{2}~\hat{Q}~e^{\frac{\gamma}{2}t}~,
\label{gran35}
\end{equation}
\begin{equation}
\hat{p}_{x,c}~=~\frac{m}{2}~\left( \dot{x}~+~\frac{\gamma}{2}~x
\right)~e^{\gamma t}~=~\frac{1}{2}~\hat{P}~e^{\frac{\gamma}{2}t}~.
\label{gran36}
\end{equation}

Inserting this into $\hat{H}_B$ (Eq, (27)) turns it into
\begin{equation}
\hat{H}_{B,c}~=~\frac{1}{m}~ \hat{p}_{x,c}~ \hat{p}_y~+~m \left(
  \omega^2 - \frac{\gamma^2}{4} \right) \hat{x}~ \hat{y}_c~=~H_{\Omega}~,
\label{gran37}
\end{equation}
since
\begin{equation}
D~=~\frac{\gamma}{2}~ (\hat{y}_c ~\hat{p}_y~-~\hat{x}~ \hat{p}_{x,c})~=~0~.
\label{gran38}
\end{equation}

$\hat{H}_{B,c}$ in (37) when expressed in terms of
$x$ and $\dot{x}$ is identical to $\hat{H}_{exp}$ as given in (19). However,
$\hat{H}_{B,c}$ is no longer a Hamiltonian that provides the correct equations
of motion, the reason being that the constraints contain an explicit time dependence.

On the other hand, $\hat{y}_c$ as defined in (35) now fulfilles the equation
of motion for $x$, i.e.,
\begin{equation}
\ddot{\hat{y}}_c~-~\gamma ~\dot{\hat{y}}_c~+~\omega^2 ~\hat{y}_c~=~\frac{1}{2}
\left( \ddot{x}~+~\gamma \dot{x}~+~\omega^2 x~\right)~e^{\gamma t}~=~0~.
\label{gran39}
\end{equation}

\subsection{ The case $a$ = 0}

For this choice of $a$ it follows that $c~=~\frac{1}{\gamma}$ and $b~=-~\frac{m}{\gamma}~\left(
  \omega^2 - \frac{\gamma^2}{4} \right)$. The canonical variables that are still missing attain in this case the values
\begin{equation}
\hat{y}_a~=~\frac{1}{\gamma}~\left( \dot{x}~+~\frac{\gamma}{2}~x
\right)~e^{\gamma t}~=~\frac{1}{m \gamma}~\hat{P}~e^{\frac{\gamma}{2}t}~,
\label{gran40}
\end{equation}

\begin{equation}
\hat{p}_{x,a}~=~-~\frac{m}{\gamma}~\left(
  \omega^2 - \frac{\gamma^2}{4} \right)~x~e^{\gamma t}~=~-~\frac{m}{\gamma}~ \Omega^2~\hat{Q}~e^{\frac{\gamma}{2}t}~. 
\label{gran41}
\end{equation}

Inserted into $\hat{H}_B$, this now yields
\begin{equation}
\hat{H}_{B,a}~=~\frac{\gamma}{2}~ (\hat{y}_a ~\hat{p}_y~-~\hat{x}~ \hat{p}_{x,a})~=~D~,
\label{gran42}
\end{equation}
with
 \begin{equation}
H_{\Omega}~=~\frac{1}{m}~ \hat{p}_{x,a}~ \hat{p}_y~+~m \left(
  \omega^2 - \frac{\gamma^2}{4} \right) \hat{x}~ \hat{y}_a~=~0~,
\label{gran43}
\end{equation}
i.e., just the opposite situation to the case $c = 0$.

Again, $\hat{H}_{B.a}$ is no longer a proper Hamiltonian function that
provides the correct equations of motion (see the comments in the previous case).

In this case, the equation of motion for $\hat{y}_a$ leads to
\begin{equation}
\ddot{\hat{y}}_a~-~\gamma \dot{\hat{y}}_a~+~\omega^2 \hat{y}_a~=~ \left[
  \frac{1}{\gamma}~\frac{d}{dt}~\left( \ddot{x}~+~\gamma \dot{x}~+~\omega^2 x~\right)~+~\frac{1}{2}
\left( \ddot{x}~+~\gamma \dot{x}~+~\omega^2 x~\right)~\right]~e^{\gamma t}~=~0~.
\label{gran44}
\end{equation}

\subsection{ The case $b$ = 0}

Now one obtains $c~=~\frac{\gamma/4}{\omega^2}$ and $a~=~\frac{m}{2 \omega^2}~\left(
  \omega^2 - \frac{\gamma^2}{4} \right)$, leading to
\begin{equation}
\hat{y}_b~=~\left( \frac{\gamma/4}{\omega^2}~\dot{x}~+~\frac{1}{2}~x
\right)~e^{\gamma t}~=~\frac{1}{2 \omega^2}~\left( \frac{\gamma}{2 m}
  \hat{P}~+~\Omega^2 \hat{Q} \right)~e^{\frac{\gamma}{2}t}~,
\label{gran45}
\end{equation}

\begin{equation}
\hat{p}_{x,b}~=~\frac{m}{2 \omega^2}~\left(
  \omega^2 - \frac{\gamma^2}{4} \right)~\dot{x}~e^{\gamma t}~=~\frac{m
  \Omega^2}{2 \omega^2}~\left( \frac{1}{m}
  \hat{P}~-~\frac{\gamma}{2} \hat{Q} \right)~e^{\frac{\gamma}{2}t}~. 
\label{gran46}
\end{equation}

Comparison with $\hat{H}_B$ (Eq. (27) or (19)) shows that now
\begin{equation}
H_{\Omega}~=~\frac{1}{m}~ \hat{p}_{x,b}~ \hat{p}_y~+~m \left(
  \omega^2 - \frac{\gamma^2}{4} \right) \hat{x}~ \hat{y}_b~=~\left(
  1~-~\frac{\gamma^2/4}{\omega^2} \right)~\hat{H}_B~=~\frac{\Omega^2}{\omega^2}~\hat{H}_B~,
\label{gran47}
\end{equation}

\begin{equation}
D~=~\frac{\gamma}{2}~ (\hat{y}_b ~\hat{p}_y~-~\hat{x}~ \hat{p}_{x,b})~=~\frac{\gamma^2/4}{\omega^2}~\hat{H}_B
\label{gran48}
\end{equation}
are valid.

The equation of motion for $\hat{y}_b$ now reads
\begin{equation}
\ddot{\hat{y}}_b~-~\gamma \dot{\hat{y}}_b~+~\omega^2 \hat{y}_b~=~ \left[
  \frac{\gamma/2}{\omega^2}~\frac{d}{dt}~\left( \ddot{x}~+~\gamma \dot{x}~+~\omega^2 x~\right)~+~\frac{1}{2}
\left( \ddot{x}~+~\gamma \dot{x}~+~\omega^2 x~\right)~\right]~e^{\gamma t}~=~0~.
\label{gran49}
\end{equation}

\section{ Conclusions}

There are several approaches for the description of dissipative systems taking
into account the environment in an effective way while still conforming to the
conventional Hamiltonian formalism. In the case of the damped harmonic
oscillator, they lead to an equation of motion for the damped system including
a linear velocity-dependent friction force. For the Bateman Hamiltonian, the
environment is substituted by one additional variable and the corresponding
momentum. It has been shown in our analysis that this variable, fulfilling a
formal equation of motion with an accelerating force, is not just a position
variable of a separate system that absorbs the energy dissipated by the damped
system. In fact, the na\"ive idea that the Bateman system represents a
degree of freedom interacting with a one-dimensional thermal bath is not quite
appropriate: the system is so intricate that the Hamiltonian is not written as
the sum of the two individual Hamiltonians plus an interaction term. However,
after imposing suitable constraints (which equate $\hat{H}_B$ and $\hat{H}_{exp}$) we arrive at
the Expanding Coordinates system that does describe the damped harmonic
oscillator, and whose variables are connected with physical position and
momentum via a non-canonical transformation.

More precisely, the link between the Bateman approach and the ones using canonical Hamiltonians with only one variable and the corresponding momentum can be achieved via an approach using an exponentially-expanding coordinate system since, in this case, the Hamiltonian is a constant of motion and can be compared with the constant Hamiltonian of the Bateman model.

It emerges that the constraints are not uniquely defined since there are more
parameters than equations for their determination. Several illustrative
examples for the choice of the constraints are discussed in detail. In
general, the Hamiltonian $\hat{H}_B$, after imposing the constraints, is no
longer a Hamiltonian in the sense that it would provide correct equations of
motion, since the constraints contain an explicit time dependence.

\begin{figure}[ht!]
\centering
\scalebox{1} 
{
\begin{pspicture}(0,0)(6,9)

\psline[linewidth=0.04]{->}(1.8,7.9)(-2.8,1.5)
\psline[linewidth=0.04]{->}(2.6,7.9)(7.2,1.5)

\psline[linewidth=0.04]{<->}(1.4,4)(-2.5,1.3)
\psline[linewidth=0.04]{<->}(3,4)(6.9,1.3)

\psline[linewidth=0.04]{<->}(-2,0.5)(6.4,0.5)

\rput[l](1.5,8.8){Bateman}
\rput[l](1.3,8.3){$(\hat x,\hat p_x,\hat y,\hat p_y)$}

\rput[l](0.7,4.8){Physical coordinates}
\rput[l](1.8,4.3){$(x,p)$}

\rput[l](-4.2,1){Expanding}
\rput[l](-4.2,0.6){coordinates}
\rput[l](-3.7,0.1){$(\hat{Q},\hat{P})$}

\rput[l](6.9,1){Caldirola-}
\rput[l](7.2,0.6){-Kanai}
\rput[l](7.3,0.1){$(\hat{x},\hat{p})$}

\rput[l](-0.5,1){time-dependent canonical transf.}
\rput[l]{35}(-1.6,1.5){non-canonical transf.}
\rput[l]{-35}(3.3,3.4){non-canonical transf.}

\rput[l]{55}(-2.2,3.1){constraint}
\rput[l]{-55}(3.7,7.9){constraint embeded in a}
\rput[l]{-55}(3.3,7.7){time-dependent canonical transformation}
\end{pspicture} 
}
\caption{This diagram shows the connections between the different canonical
descriptions and their relation to a formulation in terms of physical
position and momentum variables.}
\end{figure}
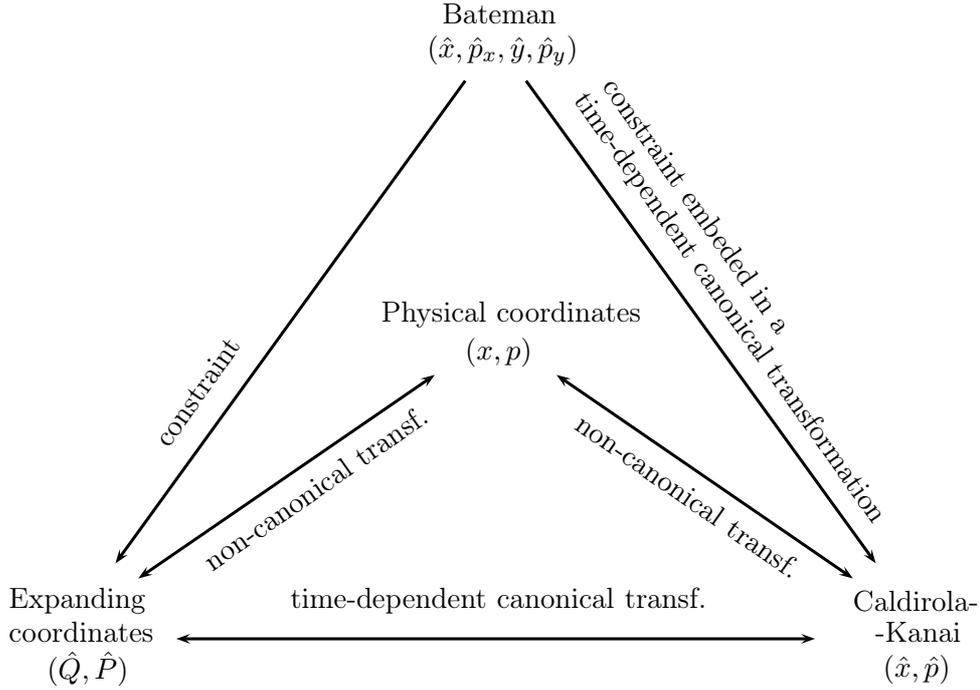

The relation between the variables of the Bateman system and the one in
expanding coordinates can be given explicitly in terms of the (physical) position and
velocity of the damped system; the connection with the CK-model can finally be
achieved via the time-dependent canonical transformation between this model and the one using
the expanding coordinates. For this purpose, $\hat{F}_2$ (see Eq. (26)) can be
written in terms of 
$\hat{Q}$ and $\hat{p}$ as 
$\hat{F}_2(\hat{Q},\hat{p},t) = \hat{Q}~ \hat{p}~ e^{- \frac{\gamma}{2}t}~+~ m  \frac{\gamma}{4}~ \hat{Q}^2$ 
with 
$
\frac{\partial}{\partial t}~\hat{F}_2(\hat{Q},\hat{p},t)~=~-~\frac{\gamma}{2}~\hat{Q}~ \hat{p}~e^{- \frac{\gamma}{2}t}~\hat{=}~\frac{\gamma}{2}~\hat{x}~\hat{p}~\hat{=}~\frac{\gamma}{2}~x ~\dot{x}~e^{\gamma t}$, 
so
$\hat{H}_{CK}~=~\hat{H}_{exp}~+\frac{\partial}{\partial t}~\hat{F}_2(\hat{Q},\hat{p},t)$ is valid. 
A different way to embed the
time-dependent constraints in a time-dependent canonical transformation to get
directly from the Bateman Hamiltonian to the CK-Hamiltonian has been shown in
\cite{14}. Both approaches to describe the dissipative system, the one in
expanding coordinates and the one by Caldirola--Kanai, can be related with a
description in terms of the physical position and momentum variables (as
expressed in Eq. (6) with the friction force) via (different) non-canonical
transformations according to Eqs. (16) and (21), respectively.

\section*{Acknowledgments}
One of the authors (D.S.) wishes to thank Professor Victor Aldaya for the
invitation to Granada that provided the basis for this work. This work was
partially supported by the Fundaci\'{o}n S\'{e}neca, the Spanish MICINN and
Junta de Andaluc\'{i}a under projects 08814/PI/08, FIS2011-29813-C02-01 and
FQM219-FQM1951, respectively.


\begin{thebibliography}{16}

\bibitem{1} Caldeira A O and Leggett A J 1981 {\it Phys. Rev. Lett.} {\bf 46} 211; Caldeira A O and Leggett A J 1983 {\it Ann. Phys. (N.Y.)} {\bf 149}
  374; {\it ibid} 1983 {\bf 153} 445(E) 
\bibitem{2} Weiss U 1993 {\it Quantum Dissipative Systems} (Singapore: World Scientific)
\bibitem{3} Bateman H 1931 {\it Phys. Rev.} {\bf 38} 815
\bibitem{4} Morse P M and Feshbach H 1953 {\it Methods of Theoretical
    Physics}, Vol. 1 (New York: McGraw-Hill)
\bibitem{5} Bopp F 1973 {\it Sitz.-Ber. Akad. Wiss., Math.-Natur. Kl.} 67
\bibitem{6} De Brito A L and Baseia B 1989 {\it Phys. Rev.} A {\bf 40} 4097
\bibitem{7} Aliaga J, Crespo G and Proto A N 1991 {\it Phys. Rev.} A {\bf 43} 595
\bibitem{8} Celeghini E, Rasetti M and Vitiello G 1992 {\it Ann. Phys.} (NY)
  {\bf 215} 156
\bibitem{9} Vitiello G 2001 {\it My Double Unveiled - The Dissipative Model of
    Brain} (Amsterdam: John Benjamin) 
\bibitem{10} Blasone M and Jizba P 2002 {\it Can. J. Phys.} {\bf 80} 645 
\bibitem{11} Blasone M and Jizba P 2004 {\it Ann. Phys.} (NY)
  {\bf 312} 354
\bibitem{12} Blasone M, Jizba P and Vitiello G 2001 {\it Phys. Lett.} A {\bf
    287} 205
\bibitem{13} 't Hooft G 1999 in: {\it Basics and Highligts of Fundamental
    Physics} Erice, hep-th/0003005 
\bibitem{14} Blasone M, Jizba P, Scardigli F and Vitiello G 2006 {\it Phys. Lett.} A {\bf
    373} 4106
\bibitem{15} Chru\'{s}ci\'{n}ski D and Jarkowski J 2006 {\it Ann. Phys.} (NY)
  {\bf 321} 854
\bibitem{16} Chru\'{s}ci\'{n}ski D and Jarkowski J 2006 {\it Ann. Phys.} (NY)
  {\bf 321} 840
\bibitem{17} Heng T-H, Lin B-S and Jing S-C 2010 {\it Chin. Phys. Lett.}  {\bf
      27} 090302
\bibitem{18} Majima H and Suzuki A 2011 {\it Ann. Phys.} (NY)
  {\bf 326} 3000
\bibitem{19} Caldirola P 1941 {\it Nuovo Cimento} {\bf 18} 393
\bibitem{20} Kanai E 1948 {\it Progr. Theor. Phys.} {\bf 3} 440
\bibitem{21} Luo T and Guo Y 2010 arXiv:0803.2330v4 [math-ph]
\bibitem{22} Baldiotti M C, Fresneda R and Gitman D M 2011 {\it Phys. Lett.} A {\bf
    375} 1630
\bibitem{23} Bender C M and Gianfreda M 2013 arXiv:1305.7107v1 [hep-th]
\bibitem{24} Schuch D 1997 {\it Phys. Rev.} A {\bf 55} 935
\bibitem{25} Pedrosa I A 1987 {\it J. Math. Phys.} {\bf 28} 2662
\bibitem{26} Schuch D 1990 {\it Int. J. Quant. Chem., Quant. Chem. Symp.} {\bf
    24} 767
\bibitem{27} Schuch D 1999 {\it Int. J. Quantum Chem.} {\bf 72} 537
\bibitem{28} Lemos N A 1979 {\it Am. J. Phys.} {\bf 47} 857
\bibitem{29} Meinert G 1996 {\it Diplom-Thesis} J. W. Goethe-Universit\"at, Frankfurt-Main 
\bibitem{30} Yu L H and Sun C P  1994 {\it Phys. Rev.} A {\bf 49} 592
\bibitem{31} Sun C P and Yu L H  1995 {\it Phys. Rev.} A {\bf 51} 1845
\bibitem{32} Guerrero J, L\'{o}pez-Ruiz F F, Aldaya V and Coss\'io F 2012 {\it J. Phys. A: Math. Theor.} {\bf 45} 475303
\bibitem{33} Cerver\'{o} J M and Villarroel J 1984 {\it J. Phys. A: Math. Gen.} {\bf 17} 1777
\bibitem{34} Dodonov V V and Man'ko V I 1979  {\it Phys. Rev.} A {\bf 20} 550
\end{thebibliography}
\end{document}